\documentclass{article}
\usepackage[english]{babel}

\usepackage[a4paper,top=2cm,bottom=2cm,left=3cm,right=3cm,marginparwidth=2cm]{geometry}

\usepackage{setspace}
\usepackage{graphicx}
\graphicspath{{./Fig/}}
\usepackage{amsmath}
\usepackage{amssymb}
\usepackage{amsthm}
\usepackage{hyperref}
\usepackage{booktabs}
\usepackage{array}
\usepackage[numbers]{natbib}
\usepackage{float}
\usepackage{verbatim}

\def\keywords#1{\par\medskip\noindent\textbf{Keywords:}~#1\par\medskip}
\providecommand{\ccsdesc}[2][]{}
\providecommand{\Description}[1]{}
\newenvironment{CCSXML}{\comment}{\endcomment}

\begin{document}

\title{Graph-Informed Deep Factor Model for Statistical Arbitrage}


\author{Evelyn Zhu$^{1}$, Mucheng Zhu$^{1}$, Brendan Martin$^{1}$, Mihai Cucuringu$^{1,2,3}$\\}

\date{
{\fontsize{11.4}{11}\selectfont
\textit{$^{1}$Department of Mathematics, University of California, Los Angeles (UCLA)\\
        $^{2}$Department of Statistics, University of Oxford\\
        $^{3}$Oxford-Man Institute of Quantitative Finance, University of Oxford}
}
}

\maketitle

%


\begin{center}
\textbf{Abstract}
\end{center}

Statistical arbitrage exploits temporary mispricings between similar assets in the cross-section of daily equity returns. We propose a nonlinear single-factor model, called a Graph-Informed Deep Factor (GIDF) model, whose per-stock loadings have cross-sectional dependence conditional on a latent network. 
The network captures cross-sectional return dependencies that remain after removing the dominant eigenmodes of the return correlation matrix, thereby providing a complementary representation of inter-asset relationships beyond the leading sources of systematic variation.
It is reconstructed from data using a novel method that we call the Factor-Adjusted Dynamic EWMA (FADE).  A signed-edge graph attention mechanism then combines the reconstructed network with exogenous firm characteristics and macroeconomic variables. This graph-informed specification is designed to produce residuals that more effectively separate idiosyncratic mean-reverting dynamics from cross-sectional co-movements, yielding economically useful signals for daily statistical arbitrage.
In a comprehensive empirical study on daily U.S. equities from 2015 to 2024, GIDF outperforms classical factor models and state-of-the-art deep learning methods, achieving an out-of-sample Sharpe ratio of $1.78$ under the canonical Avellaneda--Lee rule.


\section{Introduction}
\label{sec:intro}
Statistical arbitrage exploits temporary mispricings in the cross-section of stock returns by trading a stock against a mimicking portfolio constructed to have the same systematic risk exposure. The temporary mispricings are computed from the residual return series after systematic sources of market return variation have been removed. When the residual deviates from its mean, the arbitrageur trades against the deviation and profits if the residual subsequently mean-reverts~\cite{AvellanedaLee2010}. Our motivation is therefore to construct residuals that carry strong tradable idiosyncratic dynamics. To do this, we model the common market structures via a factor model $R_{i,t} = w_{i,t}f_{t} + \varepsilon_{i,t}$ in which the loadings $w_{i,t}$ are nonlinear functions of a latent network, firm characteristics, and macroeconomic variables. 
Crucially, the network component induces cross-sectional dependence among factor loadings. Our hypothesis is that conditioning the loadings on the local dependency structure of each asset leads to a more economically meaningful decomposition of returns and, consequently, to residuals with stronger exploitable temporal dynamics than those obtained from characteristic-only factor models.

\begin{figure}[!t]
  \includegraphics[width=1.02\textwidth]{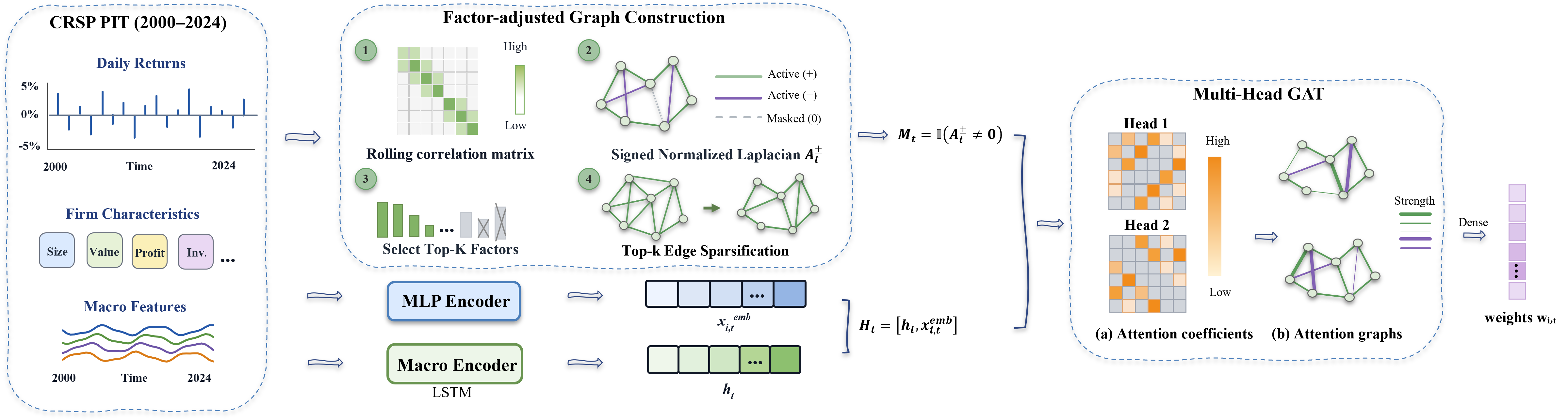}
  \caption{\normalfont The Graph-Informed Deep Factor (GIDF) pipeline. Inputs from the CRSP point-in-time universe (2000--2024) feed the FADE graph construction, producing the signed sparse adjacency $A_t$ in Eq.~\eqref{eq:fade-adj}. An edge-conditioned graph attention layer (Eqs.~\eqref{eq:egat-score}--\eqref{eq:egat-scale}) consumes the standardized signed edge weights $\tilde A_{t,ij}$ together with firm characteristics concatenated to the LSTM-encoded macroeconomic state, producing factor loadings $w_{i,t}$, which define the one-factor process $\mathcal{M}_t = 1 + \sum_{i \in \mathcal{U}_t} w_{i,t} R_{i,t}$ used as the training target (Section~\ref{sec:loss}).}
  \Description{The full pipeline for GIDF. Left: CRSP point-in-time inputs from 2000 to 2024 -- daily returns, firm characteristics, and macro features. Middle:four-step FADE graph construction: rolling EWMA-Spearman correlation, top-$K$ factor stripping, top-$K_{\text{nn}}$ signed edge sparsification, and symmetrization to produce the signed sparse adjacency $A_t$. Right: edge-conditioned multi-head graph attention (E-GAT) over the FADE graph, in which each edge weight is passed through a per-edge multiplier MLP $\phi$ before entering the attention score, combined with rank-standardized firm characteristics concatenated to the LSTM-encoded macroeconomic state, producing per-stock factor loadings.} 
  \label{fig:teaser}
\end{figure}

Classical approaches decompose the statistical arbitrage pipeline into two independent steps:  identifying a factor model, then  designing  a trading rule for its residuals. Contributions to this pipeline have come from two largely separate strands of the literature, operating primarily at monthly and daily frequencies. The first, on deep-learning asset pricing~\cite{GuKellyXiuAE2021,ChenPelgerZhu2024,GuKellyXiu2020}, relaxes the linearity of classical factor models~\cite{FamaFrench2015} but is estimated at monthly frequency and models each stock's factor loadings as a function only of its own characteristics in isolation, a convention introduced by Instrumented PCA~\cite{KellyPruittSu2019}. The second strand~\cite{GuijarroPelger2024}, which focuses on statistical arbitrage at daily frequency, retains classical PCA or Fama--French factors for the residual-construction step, placing its deep-learning contribution on the trading side rather than on the factor model. The current state of the art, Attention Factors for Statistical Arbitrage~\cite{EpsteinPelger2025b}, partially bridges this frequency gap by training a conditional latent factor model on daily data. However, its factor construction takes only firm characteristics as input to its attention mechanism, with factor loadings derived from characteristic-driven portfolio weights. Returns enter the model in two places:  as training targets and as inputs to a downstream trading policy over the resulting residuals, never as inputs to the factor-loading map itself. Our approach differs in the input to the attention layer: GIDF conditions the loadings on a signed graph derived from cross-sectional residual return correlations after factor stripping. As such, each stock's loading is a function of the residual-correlation neighborhood it sits in rather than of its own characteristics alone. The distinction is therefore not the use of attention or cross-sectional information per se, but the explicit conditioning of the attention mechanism on an estimated factor-adjusted return-dependency graph. This introduces a data-driven structural prior that complements characteristic-based cross-sectional interactions.

Two questions remain open at the intersection of these literatures. First, the role of cross-sectional dependency structure in determining factor loadings has received limited attention in the daily setting. Recent work in monthly asset pricing~\cite{ZhuGraphFactor2022} incorporates a Pearson correlation graph into loading estimation and shows that graph-based conditioning can improve the factor tangency portfolio. We address this question by developing a graph construction procedure specifically designed to improve residual quality in the daily statistical-arbitrage setting. Second, deep factor models for daily statistical arbitrage remain underexplored: monthly deep factors are not calibrated to daily mean-reversion, and daily approaches typically couple factor and trading policy in a joint objective, rendering residual quality inseparable from execution choices.

\paragraph{Main Contribution.} This paper introduces a Factor-Adjusted Dynamic EWMA (FADE) graph construction method and incorporates it into a novel nonlinear single-factor model for daily statistical arbitrage. FADE constructs a daily signed dependency graph from returns after removing pervasive market factors, thereby isolating the idiosyncratic cross-sectional co-movement that remains. We incorporate this graph into the factor model through a masked signed-edge graph attention mechanism, which conditions the factor loadings on the residual dependency structure. The resulting Graph-Informed Deep Factor (GIDF) model is trained as a residual-construction step, independent of the downstream trading rule, under an unconditional no-arbitrage moment condition~\cite{HansenJagannathan1991}. 
The training objective combines an unconditional asset-pricing moment condition, a return-reconstruction term, and a graph-based attention regularizer. Although the objective does not explicitly optimize residual mean-reversion or trading profitability, it is designed to produce economically meaningful return decompositions. Whether the resulting residuals exhibit exploitable mean-reverting dynamics is assessed independently through out-of-sample statistical and trading experiments.
As illustrated in Figure~\ref{fig:teaser}, a fixed Avellaneda--Lee s-score trading rule~\cite{AvellanedaLee2010}  applied identically to GIDF and every benchmark converts these residuals into positions.
Holding the downstream trading rule fixed provides a controlled comparison of the economic usefulness of the residuals produced by the different factor models, although differences in residual scale, trading frequency, portfolio exposures, and turnover may also contribute to the observed performance differences.
Under this framework, graph-informed residuals outperform both classical factor models and characteristic-only deep-learning benchmarks, suggesting that cross-sectional dependency structure provides a distinct source of pricing information beyond firm characteristics alone.

\section{Related Work}
\label{sec:related}

\subsection{Statistical arbitrage}
This paper builds on the statistical arbitrage literature, organized around three problems: factor decomposition, residual signal extraction, and portfolio allocation. Classical parametric methods focus on  pairwise or small-portfolio strategies based on distance, cointegration, or copula techniques. Methods that exploit large panels of stock returns include the PCA factor model of Avellaneda and Lee~\cite{AvellanedaLee2010}, which trades mean-reverting residuals through an s-score rule that has become the reference for factor-residual statistical arbitrage. More recent deep-learning approaches retain classical factors for the residual step and innovate on the trading side. Guijarro-Ordonez, Pelger, and Zanotti~\cite{GuijarroPelger2024} extract general time-series patterns from PCA residuals using a deep sequence model at daily frequency. Most closely related in application, Epstein, Wang, Choi, and Pelger~\cite{EpsteinPelger2025b} extend this framework by jointly optimizing an attention-based conditional latent factor model with a sequence-model trading policy for the objective of net-of-cost Sharpe, which conflates factor quality with policy quality. Our approach isolates the factor decomposition itself, allowing residual quality to be evaluated independently of downstream trading decisions. 

\subsection{Conditional factor models}
The workhorse models in equity asset pricing are the linear factor models exemplified by Fama and French~\cite{FamaFrench2015} and Carhart~\cite{Carhart1997}. The theoretical foundation for factor-based return decompositions comes from arbitrage pricing theory~\cite{Ross1976}, and the stochastic discount factor framework of Hansen and Jagannathan~\cite{HansenJagannathan1991} justifies treating residuals as economically meaningful; a broader treatment appears in Cochrane~\cite{Cochrane2009}. Instrumented principal component analysis~\cite{KellyPruittSu2019} extends the linear tradition by letting latent factor loadings depend linearly on observable firm characteristics. Recent deep-learning approaches generalize this mapping to nonlinear functions.
Gu, Kelly, and Xiu benchmark a broad class of machine-learning methods for return prediction~\cite{GuKellyXiu2020} and propose a conditional autoencoder whose latent factor loadings are learned as nonlinear functions of firm characteristics~\cite{GuKellyXiuAE2021}. Chen, Pelger, and Zhu~\cite{ChenPelgerZhu2024} formulate factor estimation as a stochastic discount factor (SDF) problem trained under an adversarial no-arbitrage moment condition, thereby learning a nonlinear mapping from firm characteristics to the SDF. Across these approaches, each stock’s factor loading is a function only of its own characteristics. The panel is treated as an independent collection of assets, and cross-sectional dependencies between assets enter only through the systematic factor component. Our approach instead conditions the loading function on the entire cross-sectional configuration of the other assets through attention over a data-driven dependency graph, allowing the factor structure to reflect inter-asset relationships in addition to firm-specific characteristics.

\subsection{Graph--based methods in finance}
Graph neural networks provide a natural framework for encoding cross-asset dependencies. Building on the graph convolutional network of Kipf and Welling~\cite{KipfWelling2017}, a range of studies have applied graph neural networks to financial return prediction. Chen et al.~\cite{ChenWei2018} use corporation-relationship graphs, Sawhney et al.~\cite{Sawhney2021} apply hypergraph convolutions over multi-relational networks, and Matsunaga et al.~\cite{Matsunaga2019} develop a rolling-window graph learning framework. Within this return-prediction thread, signed graph attention over raw correlation networks has been applied to next-period return forecasting~\cite{DYGAP2023}, a task distinct from ours. In parallel, a separate thread applies graph-based methods  specifically to statistical arbitrage at the trading or clustering layer, using signed-graph clustering machinery developed by Cucuringu et al.~\cite{Cucuringu2020}. Jin, Cucuringu, and Cartea~\cite{CarteaCucuringuJin2023} cluster the residual-correlation matrix into arbitrage portfolios, Bennett, Cucuringu, and Reinert~\cite{BennettCucuringu2022} incorporate lead-lag structure into network clustering for multivariate time series, and Korniejczuk and {\'S}lepaczuk~\cite{Ollikainen2024} extend this direction to multi-pair strategies. In both threads, the graph is used for downstream trading, clustering, or prediction. In contrast, our approach incorporates graph structure directly into the factor-model loadings, using a signed graph derived from residual correlations after factor stripping to condition each stock's factor loading
on the residual dependency structure. 

\section{Methodology}
\label{sec:method}

\subsection{Problem setup}
\label{sec:problem}
We consider a universe of $N$ assets with daily returns $R_t \in \mathbb{R}^N$, $d_x$-dimensional time-varying firm characteristics $x_{i,t} \in \mathbb{R}^{d_x}$, and a $d_m$-dimensional macroeconomic state $m_t \in \mathbb{R}^{d_m}$   over trading days  $t = 1, \dots, T$. The point-in-time tradable universe on day $t$ is $\mathcal{U}_t \subseteq \{1, \dots, N\}$.

A statistical arbitrage strategy trades the residuals of a  systematic-plus-idiosyncratic decomposition $R_{i,t} = \chi_{i,t} + \varepsilon_{i,t}$, where $\chi_{i,t} = w_{i,t} f_t$ is the systematic component. The Avellaneda--Lee trading rule is applied independently to each residual series $(\varepsilon_{i,t})_{t \in [T]}$, so strategy  performance depends on how strongly each $\varepsilon_{i,t}$ mean-reverts. A $K=1$ factor model will not, in general, deliver a diagonal residual covariance. Rather than targeting cross-sectional independence, we aim for stronger univariate mean-reversion than a characteristic-only factor model produces.

Conditioning the factor-loading function on a signed dependency graph allows the direction of the estimated systematic component to adapt to the prevailing cross-sectional configuration of the market. However, conditional on the information available at a given time, the resulting systematic return component remains rank-one. Consequently, the graph does not generally eliminate multiple independent sources of local co-movement. Our empirical hypothesis is that it nevertheless improves the choice of the factor direction and the economic usefulness of the resulting residuals.

\subsection{FADE graph construction}
\label{sec:fade}

For each trading day, we construct a signed sparse dependency graph from a rolling 252-day window of daily returns.

\paragraph{EWMA Spearman correlation.} We compute an EWMA-weighted Spearman rank correlation matrix $\hat C_t \in \mathbb{R}^{N \times N}$ with halflife 60 trading days, so that recent observations dominate the estimate and the graph
adapts to regime changes~\cite{PafkaKondor2003,FosterNelson1996}. 
For reproducibility, we distinguish between the rank transformation and the subsequent exponential weighting. Let $\mathcal{W}_t=\{t-L,\ldots,t-1\}$ denote the historical estimation window, with $L=252$, and define normalized exponential weights by
\[
\omega_{t,s}=
\frac{2^{-(t-1-s)/h}}
{\sum_{u\in\mathcal{W}_t}2^{-(t-1-u)/h}},
\qquad s\in\mathcal{W}_t,
\]
where $h=60$ denotes the half-life in trading days. We compute the exponentially weighted correlation of the rank-transformed return series.
Concretely, we compute ordinary within-window ranks on the raw return panel, with ties broken by averaging. The exponential weights $\omega_{t,s}$ are then applied as a per-row scaling $\sqrt{\omega_{t,s}}$ of the rank matrix, and Ledoit--Wolf shrinkage is subsequently fit on this weighted rank matrix. The construction is therefore equivalent to an exponentially weighted Pearson correlation of ordinary within-window ranks. Missing observations within the 252-day window are imputed by the within-window column mean after requiring each stock to have at least 50\% of its entries present; stocks below this threshold are excluded from the day-$t$ active universe, and days with fewer than 50 active stocks fall back to the identity Laplacian.
Exponential weighting reduces the effective sample size relative to the nominal estimation-window length. For normalized exponential weights $\omega_{t,s}$, the effective sample size is
\[ T_{\mathrm{eff}} = \frac{1}{\sum_{s=t-L}^{t-1}\omega_{t,s}^{2}}. \]
For our estimation window of $L=252$ trading days and a half-life of 60 trading days, the effective sample size is approximately 155 observations. Consequently, the graph estimator operates in a particularly high-dimensional regime, further motivating regularization of the estimated dependency matrix.
We use Spearman rank correlation because daily returns are heavy-tailed and rank correlation is robust to extreme values without requiring winsorization at the correlation step. Because the number of assets exceeds the effective number of historical observations, the empirical dependency matrix is potentially singular and its eigenstructure may be poorly estimated. We therefore regularize the rank-based dependency estimator using covariance shrinkage followed by rescaling to unit diagonal.
We use the standard sklearn.covariance.LedoitWolf implementation~\cite{LedoitWolf2004}, whose shrinkage target is a diagonal matrix with equal entries equal to the trace mean of the sample covariance and whose intensity is the Ledoit--Wolf 2004 closed-form oracle for linear shrinkage toward that target. We apply this estimator as a regularization heuristic on the exponentially weighted rank-transformed data; we do not re-derive its oracle intensity for the weighted-rank setting. The resulting shrunk covariance is rescaled to unit diagonal to recover a full-rank correlation matrix.
In the high-dimensional regime $N \gtrsim T$, the raw sample correlation matrix is rank-deficient; we therefore apply Ledoit--Wolf shrinkage~\cite{LedoitWolf2004}
followed by rescaling to obtain a full-rank correlation matrix.

\paragraph{Factor stripping.} We eigendecompose $\hat C_t$ into eigenvalues $\lambda_1 \ge \dots \ge \lambda_N$ with eigenvectors $v_1, \dots, v_N$. The number of dominant factors to remove is determined using  the Ahn--Horenstein eigenvalue-ratio test~\cite{AhnHorenstein2013}, floored at $K_{\min} = 5$. 
The lower bound $K_{\min}=5$ is introduced to prevent the eigenvalue-ratio criterion from retaining an excessively small number of dominant components, particularly when the leading market eigenvalue is well separated from the remaining spectrum. This choice is intended to remove a broader set of pervasive statistical modes before constructing the residual dependency graph.
\begin{equation}
\hat\Theta_t = \hat C_t -
    \sum_{k=1}^{K_t} \lambda_k v_k v_k^\top,
    \quad K_t = \max\!\big(K^{\text{ER}}_t,\, K_{\min}\big).
\label{eq:fade-strip}
\end{equation}
The residual matrix $\hat\Theta_t$ summarizes cross-sectional dependencies not explained by the top-$K_t$ eigenmodes. 
The removed eigenmodes need not correspond one-to-one to economically interpretable risk factors. Accordingly, factor adjustment in FADE refers to the removal of dominant statistical modes rather than complete neutrality with respect to a prespecified economic factor model.
Although the original matrix is a correlation matrix, the factor-adjusted matrix $\hat\Theta_t$ does not generally have unit diagonal. Its off-diagonal entries therefore represent unnormalized residual dependencies in rank space, reflecting both the strength of residual correlations and heterogeneity in residual variances. By retaining these unnormalized entries in the graph construction, FADE preserves information about the magnitude of residual variation remaining after the removal of the dominant eigenmodes.

\paragraph{Top-$K_{\text{nn}}$ sparsification.} We retain the top $K_{\text{nn}} = 20$   nearest-neighbor entries per row of $\hat\Theta_t$ by absolute magnitude and preserve their signs, obtaining a row-sparse matrix $\tilde\Theta_t$ that is generically asymmetric. Symmetrizing gives the signed sparse adjacency
\begin{equation}
A_t = \tfrac{1}{2}\big(\tilde\Theta_t + \tilde\Theta_t^\top\big).
\label{eq:fade-adj}
\end{equation}
This symmetrization corresponds to a union nearest-neighbor graph with stronger weights assigned to mutually selected neighbors. In particular, an edge selected in only one direction receives half the weight of an edge selected in both directions, holding its original dependence magnitude fixed. The resulting graph is symmetric but need not have constant node degree.
Preserving edge signs distinguishes positive from negative contemporaneous residual associations. These associations describe the direction of historical co-movement and do not, by themselves, imply convergence, divergence, cointegration, or predictive lead--lag relationships. The downstream attention layer uses both the graph support and the signed edge weights.

\subsection{Graph-informed deep factor model}
\label{sec:model}
GIDF passes firm characteristics through an edge-conditioned graph attention layer over the FADE-selected edges, augmented with the encoded macroeconomic state, and outputs a per-stock factor loading $w_{i,t}$.

\paragraph{Macro encoder.} A single-layer LSTM compresses the $d_m = 178$-dimensional macro state $m_t$ into a $d^{\text{macro}} = 4$-dimensional hidden state $\tilde m_t$, letting factor loadings adapt to slow-moving macroeconomic regimes. For each active stock $i \in \mathcal{U}_t$ we form the input embedding by concatenating firm characteristics with the shared macro state, $z_{i,t} = [x_{i,t} \parallel \tilde m_t] \in \mathbb{R}^{F_{\text{in}}}$ with $F_{\text{in}} = d_x + d^{\text{macro}} = 32$.

\paragraph{Edge-conditioned graph attention (E-GAT)} A single E-GAT layer with $H = 2$ heads and dimension $d_h = 16$  maps $Z_t = [z_{i,t}]_{i \in \mathcal{U}_t}$ to a graph-aggregated representation. For each head $h \in \{1,\dots,H\}$ the layer has independent query, key, and value projections and an attention vector,
\begin{equation}
W_q^h,\; W_k^h,\; W_v^h \in \mathbb{R}^{d_h \times F_{\text{in}}}, \qquad a^h = [a_L^h \parallel a_R^h] \in \mathbb{R}^{2d_h},
\end{equation}
where $a_L^h, a_R^h \in \mathbb{R}^{d_h}$ are the left (query-side) and right (key-side) halves of the additive attention vector $a^h$. For each active stock we form per-head queries, keys, and value vectors
\begin{equation}
q_{i,t}^h = W_q^h z_{i,t}, \quad k_{j,t}^h = W_k^h z_{j,t}, \quad V_{j,t}^h = W_v^h z_{j,t}.
\end{equation}
The pre-softmax score for edge $(i,j)$ combines an additive graph attention network (GAT) style content term with a per-edge FADE bias.
\begin{equation}
e^h_{ij,t} = \operatorname{LeakyReLU}\!\big( (a_L^h)^\top q_{i,t}^h + (a_R^h)^\top k_{j,t}^h \big) + \lambda_h \, \phi(\tilde A_{t,ij}) \, \tilde A_{t,ij},
\label{eq:egat-score}
\end{equation}
where the LeakyReLU uses negative slope $0.2$. 
The learnable per-head scalar $\lambda_h \in \mathbb{R}$ biases the FADE edge signal and is warm-started at $1.0$ so the edge term carries weight from the first training step. The signed edge weight $\tilde A_{t,ij} = A_{t,ij} / \sigma_t^A$ is standardized by the per-day scale $\sigma_t^A$
\begin{equation}
\sigma_t^A = \Big(|E_t|^{-1} \!\!\sum_{(i,j) \in E_t}\! A_{t,ij}^2\Big)^{1/2}, \quad E_t = \{(i,j) : A_{t,ij} \ne 0\},
\label{eq:egat-scale}
\end{equation}
so that $\lambda_h \tilde A$ has magnitude comparable to the content-based score in the softmax competition, and $\phi: \mathbb{R} \to \mathbb{R}$ is a per-edge multiplier learned by a multilayer perceptron (MLP) that is shared across heads. The final layer of $\phi$ is initialized with zero weights and unit bias so that $\phi \equiv 1$ at initialization and the layer reduces to a standard GAT with a signed edge bias. 
Through training, $\phi$ learns a nonlinear transformation of the FADE edge weights, allowing the model to modify their influence on the attention distribution according to the training objective. Unless positivity or monotonicity constraints are imposed on $\phi$ and $\lambda_h$, the learned transformation need not preserve the sign or ordering of the original edge weights.

Attention is softmax-normalized over the FADE-selected neighborhood $\mathcal{N}_t(i) = \{j : A_{t,ij} \ne 0\} \cup \{i\}$, with the self-loop retained for numerical stability, and per-head outputs are concatenated and linearly projected back to dimension $F_{\text{out}} = 32$:
\begin{align}
\alpha^h_{ij,t} &= \frac{\exp(e^h_{ij,t})}{\sum_{\ell \in \mathcal{N}_t(i)} \exp(e^h_{i\ell,t})}, \qquad
g^h_{i,t} = \sum_{j \in \mathcal{N}_t(i)} \alpha^h_{ij,t} \, V_{j,t}^h,
\label{eq:egat-agg}\\
g_{i,t} &= W_o \big[ g^1_{i,t} \parallel g^2_{i,t} \big], \qquad W_o \in \mathbb{R}^{F_{\text{out}} \times H d_h}.
\label{eq:egat-out}
\end{align}
Although the FADE graph contains signed edges, the softmax normalization ensures that the attention coefficients $\alpha^h_{ij,t}$ are nonnegative. The edge signs therefore influence the attention scores and the relative importance assigned to neighboring nodes, rather than directly determining the sign of their contributions to the aggregated representation. In this sense, the architecture implements sign-conditioned attention rather than explicitly signed message passing.

\paragraph{Dense head.} We concatenate the input embedding with the graph-aggregated representation, $u_{i,t} = [z_{i,t} \parallel g_{i,t}] \in \mathbb{R}^{F_{\text{in}} + F_{\text{out}}}$, and pass it through two 
ReLU-activated fully-connected layers of width $64$ with dropout $0.7$, followed by a linear projection to the scalar weight $w_{i,t}$. The implied daily one-factor process is $\mathcal{M}_t = 1 + \sum_{i \in \mathcal{U}_t} w_{i,t} R_{i,t}$. 
The vector $w_t$ parameterizes the portfolio component of the stochastic discount factor. In the residual-construction step, we additionally use the same vector to define the direction of the rank-one return projection. This constitutes a modeling restriction: in a general factor model, stochastic discount factor portfolio weights and return-factor loadings are distinct objects and need not coincide.

\subsection{Loss function and training}
\label{sec:loss}

We train GIDF using a daily-frequency approximation to the unconditional no-arbitrage moment condition of Hansen and Jagannathan~\cite{HansenJagannathan1991}, given by $\mathbb{E}[R_{i,t} \mathcal{M}_t] = 0$ per stock. The one-factor process $\mathcal{M}_t = 1 + \sum_{i \in \mathcal{U}_t} w_{i,t} R_{i,t}$ defined in Section~\ref{sec:model} follows the \citeauthor{ChenPelgerZhu2024}~ \cite{ChenPelgerZhu2024} parameterization of the pricing kernel. We train the loadings under the Hansen--Jagannathan moment condition~ \cite{HansenJagannathan1991,Cochrane2009}, which selects the loading map whose implied one-factor projection minimizes empirical pricing errors on the training panel. The resulting per-stock residuals feed the Avellaneda--Lee trading rule in Section~\ref{sec:protocol}.
The exact no-arbitrage moment condition is defined in terms of excess returns, \[ \mathbb{E}\left[(R_{i,t}-R_t^f)\mathcal{M}_t\right]=0, \] where $R_{i,t}$ denotes the raw asset return and $R_t^f$ the daily risk-free return. At daily frequency, we adopt the approximation $R_t^f \approx 0$ and accordingly use the moment condition $\mathbb{E}[R_{i,t}\mathcal{M}_t]=0$ throughout. While the daily risk-free return is small relative to typical daily return fluctuations, the resulting approximation error in the pricing moments, $\mathbb{E}[R_t^f\mathcal{M}_t]$, need not be negligible relative to expected daily returns, particularly during periods of elevated interest rates.
Denote by $\bar\mu_i$ the empirical mean of $R_{i,t} \mathcal{M}_t$ over the days on which stock $i$ is active. The no-arbitrage loss is
\begin{equation}
\mathcal{L}_{\text{noarb}} = 252 \cdot \frac{1}{N} \sum_{i=1}^N \bar\mu_i^2, \quad \bar\mu_i = \frac{1}{T_i} \sum_{t \,:\, i \in \mathcal{U}_t} R_{i,t} \mathcal{M}_t,
\label{eq:loss-noarb}
\end{equation}
where $T_i$ is the number of active days for stock $i$. Returns are winsorized at the $0.1\%$ and $99.9\%$ quantiles to prevent a small number of extreme daily returns from dominating the empirical mean and destabilizing the gradient.

This objective encourages the loadings to satisfy the no-arbitrage pricing condition but leaves residual variance unconstrained: a factor model can satisfy the no-arbitrage condition while producing residuals whose variance is close to the raw return variance, which is the failure mode that hurts downstream mean-reversion trading. We augment the objective with a residual reconstruction loss. Writing $\langle \cdot \rangle_i^t = |\mathcal{U}_t|^{-1} \sum_{i \in \mathcal{U}_t}(\cdot)$ for the cross-sectional average on day $t$,
\begin{equation}
\mathcal{L}_{\text{res}} = \frac{T^{-1}\sum_{t=1}^{T}\langle \varepsilon_{i,t}^2 \rangle_i^t}{T^{-1}\sum_{t=1}^{T}\langle R_{i,t}^2 \rangle_i^t}, \quad \varepsilon_{i,t} = R_{i,t} - f_t\, w_{i,t}, \quad f_t = \frac{\sum_{i \in \mathcal{U}_t} R_{i,t} w_{i,t}}{\sum_{i \in \mathcal{U}_t} w_{i,t}^2 + \delta},
\label{eq:loss-res}
\end{equation}
where $f_t$ is the implied one-factor return and $\varepsilon_{i,t}$ is the residual for stock $i$ on day $t$. Each day is weighted equally (via the inner cross-sectional average), preventing days with more active stocks from receiving greater weight solely because of universe size. The ratio measures the fraction of the uncentered return second moment left unexplained by the implied one-factor reconstruction. For nonnegative numerical regularization $\delta$, the reconstruction loss is bounded in $[0,1]$, provided that its denominator is positive. In the absence of numerical regularization, the reconstruction is invariant to any nonzero scalar rescaling of the loading vector $w_t$. The reconstruction objective therefore identifies the direction, but not the scale, of the loading vector. In contrast, the no-arbitrage objective is generally sensitive to the scale of $w_t$, so the two objectives play complementary roles in the estimation of the factor model.
The total training loss is
\begin{equation}
\mathcal{L} = \mathcal{L}_{\text{noarb}} + \lambda_{\text{res}} \mathcal{L}_{\text{res}} + \lambda_{\text{att}} \mathcal{L}_{\text{att}},
\label{eq:loss-total}
\end{equation}
with $\lambda_{\text{res}} = \lambda_{\text{att}} = 0.3$. The attention-supervision term $\mathcal{L}_{\text{att}} = H^{-1}\sum_{h=1}^{H}(T\,|\mathcal{U}_t|)^{-1} \sum_{t,i} \mathrm{KL}\big(p_{i,t} \,\|\, \alpha^{h}_{i,\cdot,t}\big)$ pulls each stock's per-head attention distribution $\alpha^{h}_{i,\cdot,t} = (\alpha^{h}_{ij,t})_{j \in \mathcal{N}_t(i)}$ toward the FADE prior $p_{i,t} = \operatorname{softmax}(|\tilde A_{t,ij}|)_{j \in \mathcal{N}_t(i)}$, biasing attention toward neighbors with stronger standardized edge weight; the pricing objective stays primary and the other two act as regularizers.

Neither the pricing moment restrictions nor the reconstruction term explicitly impose stationarity, negative serial dependence, or mean-reversion on the residual process. The attention regularizer likewise constrains the use of graph information rather than residual temporal dynamics. Consequently, stronger residual mean-reversion is an empirical hypothesis which we test out of sample, rather than a mathematical property guaranteed by the training objective.

\paragraph{Training.} We optimize with Adam at learning rate $10^{-4}$ and weight decay $10^{-5}$, with dropout $0.7$ in the dense head. We train for $960$ epochs with the full loss. Hyperparameters were selected on the first window's validation split and held fixed across all five windows, so the reported out-of-sample (OOS) Sharpe is not tuned per window. The full five-window training schedule takes approximately $5$ hours on a single NVIDIA RTX $5080$ GPU.
\section{Empirical Analysis}
\label{sec:empirical}

\subsection{Data}
\label{sec:data}

We use daily equity return data from the Center for Research in Security Prices (CRSP), covering January 2000 through December 2024 for a total of 6{,}289 trading days.

\paragraph{Universe.} Our universe consists of the historical S\&P 500 constituents drawn from the CRSP historical index membership table. Each PERMNO (CRSP permanent stock identifier) enters the universe on its S\&P 500 entry date and exits on its removal or delisting date, so membership is strictly point-in-time, ensuring that no forward-looking information leaks into eligibility. Delisted stocks remain in the universe up to and including their final tradable day, with the day's ordinary return $R^{\text{raw}}_{i,t}$ compounded with the CRSP delisting return $\delta_{i,t}$ via the standard convention $R_{i,t} = (1 + R^{\text{raw}}_{i,t})(1 + \delta_{i,t}) - 1$, yielding the delisting-adjusted return $R_{i,t}$ used throughout the model. 
Over 2000--2024 the universe contains 1{,}069 unique PERMNOs, with a median of 500 tradable names per trading day.

\paragraph{Characteristics.} For each stock and each trading day we compute a 28-dimensional characteristic vector $x_{i,t}$ listed in Table~\ref{tab:features}. All rolling features on day $t$ use returns and prices observed on day $t-1$ or earlier.
Each characteristic is rank-standardized: on each day $t$, each stock's characteristic value is replaced by its cross-sectional rank across the active universe $\mathcal{U}_t$, and this rank is then mapped to $[-1, +1]$ via $2\, \text{rank}/(|\mathcal{U}_t|+1) - 1$, which controls for outliers and puts all features on a common scale.

\paragraph{Macroeconomic state.} We complement the firm-level characteristics with a 178-dimensional macroeconomic state vector $m_t$ from the Federal Reserve Economic Data monthly (FRED-MD) database, covering industrial production, employment, credit, and market-based indicators.

\begin{table}[!ht]
\centering
\small
\begin{tabular}{@{}ll@{}}
\toprule
\multicolumn{2}{@{}l}{\textbf{Returns (7)}} \\
$R_k,\ k \in \{1, 5, 21, 63, 126\}$ & $k$-day cumulative return \\
$\text{mom}_{12,1}$ & 12-month minus 1-month momentum~\cite{JegadeeshTitman1993} \\
$\text{rev}_{21,5}$ & 5-day short-term reversal~\cite{LoMacKinlay1990} \\
\addlinespace
\multicolumn{2}{@{}l}{\textbf{Volatility (6)}} \\
$\text{vol}_{20},\ \text{vol}_{60}$ & Rolling return volatility \\
$\text{skew}_{60},\ \text{kurt}_{60}$ & Rolling skewness, kurtosis \\
$\text{idio\_vol}_{60}$ & Idiosyncratic vol vs.\ market \\
$\text{max\_dd}_{60}$ & Maximum drawdown \\
\addlinespace
\multicolumn{2}{@{}l}{\textbf{Liquidity (5)}} \\
$\log(\text{shvol}),\ \log(\text{dvol})$ & Log share and dollar volume \\
$\text{turnover}_{20}$ & Rolling turnover \\
Amihud$_{20}$ & Amihud illiquidity measure \\
$\text{zero\_freq}_{20}$ & Frequency of zero-return days \\
\addlinespace
\multicolumn{2}{@{}l}{\textbf{Microstructure (2)}} \\
opcl & Intraday open-to-close return \\
overnight & Overnight return \\
\addlinespace
\multicolumn{2}{@{}l}{\textbf{Size and price (4)}} \\
$\log(\text{mcap})$ & Log market capitalization \\
$\log(\text{price})$ & Log share price \\
$d_{\text{52w-hi}},\ d_{\text{52w-lo}}$ & Distance to 52-week high, low \\
\addlinespace
\multicolumn{2}{@{}l}{\textbf{Market beta (3)}} \\
$\beta_{60},\ \beta_{252}$ & Rolling market beta \\
$\rho_{60}$ & 60-day correlation with market \\
\addlinespace
\multicolumn{2}{@{}l}{\textbf{Cross-section (1)}} \\
rank & Same-day return rank in universe \\
\bottomrule
\end{tabular}
\smallskip
\caption{\normalfont Firm characteristics grouped by category.}
\label{tab:features}
\end{table}

\subsection{Evaluation setup}
\label{sec:protocol}
We use the same asset universe, rolling-window schedule, checkpoint-selection rule, and trading strategy for all models. The five expanding windows W1--W5 cover 2015--2024. In each window, the model is trained on all data available up to the year before validation, validated on the following year, and evaluated on the subsequent two-year out-of-sample test period. We select the checkpoint with the lowest validation loss and apply it unchanged to the corresponding test set.

For each model, residuals are converted into positions using a standardized-residual reversal strategy inspired by Avellaneda and Lee. The signal is constructed by normalizing the most recently available residual return by its rolling 60-day standard deviation. Positions are opened and closed using entry and exit thresholds with hysteresis. The same signal construction and threshold rule are applied to all factor models.

The trading timeline follows the canonical Avellaneda--Lee convention. At the close of each trading day $t$, the $s$-score $s_{i,t}$ is formed from the residual history $\{\varepsilon_{i,s}\}_{s \le t}$ through close $t$. Positions are entered at close $t$ based on $s_{i,t}$, and the resulting P\&L is realized over the close-$t$-to-close-$t{+}1$ return. There is no additional trading-day delay between signal formation and position entry: in practice this corresponds to execution in the closing auction, or equivalently near-close around 15{:}50 ET. The GIDF loading map itself depends on firm characteristics $x_{i,t-1}$ observed at the prior close and the contemporaneous macroeconomic state $m_t$, so no forward-looking firm-specific information enters the systematic component.

Portfolio returns from the five test periods are concatenated chronologically into a single out-of-sample daily portfolio-return series $\{R^{\text{p}}_t\}_{t=1}^{T}$, with sample mean $\bar R^{\text{p}}$, standard deviation $\sigma$, downside deviation $\sigma^{-} = \sqrt{T_{-}^{-1}\sum_{t:\,R^{\text{p}}_t<0} (R^{\text{p}}_t)^{2}}$ where $T_{-}=|\{t:R^{\text{p}}_t<0\}|$, and cumulative wealth $C_t = \prod_{s \le t}(1 + R^{\text{p}}_s)$. We report annualized Sharpe $\sqrt{252}\,\bar R^{\text{p}}/\sigma$, Sortino $\sqrt{252}\,\bar R^{\text{p}}/\sigma^{-}$, return $252\,\bar R^{\text{p}}$, volatility $\sqrt{252}\,\sigma$, maximum drawdown $\mathrm{MaxDD} = \min_t (C_t / \max_{s \le t} C_s - 1)$, and Calmar $252\,\bar R^{\text{p}}/|\mathrm{MaxDD}|$.


\subsection{Results}
\label{sec:results}

\begin{table}[!t]
\centering
\footnotesize                                 
\setlength{\tabcolsep}{4pt}                   
\begin{tabular}{@{}lrrrrrr@{}}
\toprule
Factor model & Sharpe & Sortino & MaxDD & Calmar & AnnRet & AnnVol \\
\midrule
Avellaneda--Lee PCA ($K=40$)~\cite{AvellanedaLee2010} & 0.550 & 0.552 & $-24.8\%$ & 0.378 & $9.4\%$ & $17.1\%$ \\
Factor analysis ($K=20$)~\cite{Bartholomew2011} & 0.524 & 0.519 & $-26.3\%$ & 0.342 & $9.0\%$ & $17.1\%$ \\
Adversarial SDF network ($K=1$)~\cite{ChenPelgerZhu2024} & 0.789 & 0.790 & $-20.1\%$ & 0.615 & $12.3\%$ & $15.6\%$ \\
Fama--French six-factor~\cite{FamaFrench2015,Carhart1997} & 0.930 & 0.940 & $-16.9\%$ & 0.524 & $8.9\%$ & $9.5\%$ \\
Conditional autoencoder (CAE, $K=5$)~\cite{GuKellyXiuAE2021} & 1.032 & 1.125 & $-16.4\%$ & 0.773 & $12.7\%$ & $12.3\%$ \\
\midrule
GIDF (identity) & 1.593 & 1.735 & $-7.0\%$ & 1.913 & $13.4\%$ & $8.4\%$ \\
\textbf{GIDF ($K=1$)} & \textbf{1.781} & \textbf{1.999} & $\mathbf{-7.8\%}$ & \textbf{2.008} & $\mathbf{15.6\%}$ & $\mathbf{8.7\%}$ \\
\bottomrule
\end{tabular}
\smallskip
\caption{\normalfont Out-of-sample residual-arbitrage performance for GIDF and its baselines under the canonical Avellaneda--Lee rule~\cite{AvellanedaLee2010} 
obtained by stitching together the five expanding-window out-of-sample test periods
on the CRSP point-in-time active universe ($\sim$1069 stocks, 2015--2024). $K$ denotes the number of latent factors each model outputs. All metrics annualized, gross of transaction costs.}
\label{tab:main}
\end{table}

\paragraph{Headline results.} Table~\ref{tab:main} reports the main out-of-sample results. GIDF at $K=1$ attains an annualized Sharpe ratio of $1.781$, exceeding every baseline in the table. We compare against five baselines: Avellaneda--Lee PCA at $K=40$~\cite{AvellanedaLee2010}, factor analysis at $K=20$~\cite{Bartholomew2011}, the adversarial SDF network~\cite{ChenPelgerZhu2024}, the Fama--French six-factor portfolio~\cite{FamaFrench2015,Carhart1997}, and the conditional autoencoder at $K=5$~\cite{GuKellyXiuAE2021}.
Their Sharpe ratios span $0.52$ to $1.03$, so GIDF's lead over the best is $0.749$. The two deep conditional benchmarks were built for monthly data and are retrained here at daily frequency, so their absolute levels should be read with that in mind. The comparison still holds every input fixed: the same universe, features, windows, and trading rule. 
The performance improvement relative to the external benchmarks reflects the combined effects of differences in model architecture, training objective, macroeconomic conditioning, and graph information. The incremental contribution of FADE and its associated attention supervision is assessed separately through the identity-graph ablation, which provides a more closely matched comparison.
GIDF's edge is as much about risk as return. 
GIDF achieves a higher annualized return than the external benchmarks while maintaining lower annualized volatility and smaller maximum drawdown. Its volatility reduction is particularly pronounced relative to the PCA, factor-analysis, and adversarial SDF benchmarks.
The risk-adjusted metrics follow: Sortino $1.999$ nearly doubles the best baseline's $1.125$ (CAE), and Calmar $2.008$ is more than 2.5 times the best baseline's. No baseline matches GIDF across all three principal risk-adjusted metrics.

\paragraph{Role of graph information.} A controlled ablation isolates the contribution of graph information specifically. GIDF (identity) replaces the FADE adjacency with the identity matrix, so the graph-attention layer sees only self-loops. Every other element (LSTM macro encoder, edge-conditioned attention skeleton, dense head, no-arbitrage plus residual plus attention-supervision training loss, Avellaneda--Lee trading rule) is held fixed. The identity ablation is therefore the correct control for the FADE graph itself. 

The Sharpe difference between the identity and FADE variants measures the incremental benefit of incorporating cross-stock graph information and its associated attention supervision within an otherwise matched modeling framework.

Even without any graph information, GIDF (identity) attains Sharpe $1.593$ and Sortino $1.735$, already exceeding every classical and deep baseline in Table~\ref{tab:main}. This shows the GIDF architecture is itself a strong characteristic-only deep factor model. The LSTM macro encoder, the edge-conditioned attention skeleton (which remains active on self-loops), and the HJ-plus-residual training objective together produce competitive residuals before any cross-stock structure is introduced. The FADE graph then maximizes the benefit of this architecture by giving the attention layer the exact cross-sectional residual-dependency signal it was designed to exploit.

The specific contribution of the FADE graph is isolated by the GIDF versus GIDF (identity) comparison. Adding the FADE graph lifts Sharpe from $1.593$ to $1.781$ ($\Delta = +0.188$), Sortino from $1.735$ to $1.999$, and Calmar from $1.913$ to $2.008$. Table~\ref{tab:al-sweep} shows via HAC delta-method and studentized block-bootstrap tests that this lift is statistically distinct from zero at $p<0.05$ at three of five canonical entry thresholds and at $p<0.10$ at all five, so the FADE contribution is both economically meaningful and statistically robust across the trading-rule grid. Two aspects of the FADE construction help explain this lift.
First, because FADE strips the market factors before building edges, its edges encode idiosyncratic co-movement rather than shared market exposure: when sizing a stock's contribution to $\mathcal{M}_t$, the loading can pool information across the stock's residual neighbors instead of treating each residual in isolation. Second, FADE keeps edge signs, so the loading separates neighbors that co-move (convergent) from those that move oppositely (divergent), a distinction a sign-blind graph collapses into one blurred relationship. Both effects act at the loading step, shaping the residuals subsequently traded by the univariate Avellaneda--Lee rule.

\paragraph{Hyperparameter sensitivity.} We vary each of FADE's construction parameters one at a time around its headline value, holding every other element of GIDF (model architecture, trading rule, universe, training schedule) fixed. Table~\ref{tab:fade-sensitivity} reports stitched out-of-sample performance for each variant. All five evaluated variants fall within $-0.174$ Sharpe of the headline. Two interpretable patterns emerge. First, longer EWMA halflife monotonically degrades performance ($1.784 \to 1.781 \to 1.607$ at $h \in \{30, 60, 120\}$ days), consistent with FADE's role in tracking regime shifts rather than long-run co-movement. Second, over-stripping the factor structure ($K_{\min}=15$) costs $-0.095$ Sharpe relative to the headline $K_{\min}=5$. The $K_{\text{nn}}=10$ variant retains most of the lift at half the edge density ($-0.068$).

\begin{table}[!t]
\centering
\small
\begin{tabular}{@{}llrrrrr@{}}
\toprule
Dimension & Variant & Sharpe & $\Delta$Sharpe & Sortino & MaxDD & Calmar \\
\midrule
Headline       & $K_{\text{nn}}$=20, $K_{\min}$=5, $h$=60 & \textbf{1.781} & \textbf{---}      & \textbf{1.999} & $\mathbf{-7.8\%}$ & \textbf{2.008} \\
\midrule
Top-$K_{\text{nn}}$      & $K_{\text{nn}}$=10        & 1.713 & $-0.068$ & 1.897 & $-7.0\%$ & 2.100 \\
Factor floor   & $K_{\min}$=8                        & 1.729 & $-0.052$ & 1.878 & $-8.6\%$ & 1.814 \\
               & $K_{\min}$=15                       & 1.686 & $-0.095$ & 1.815 & $-8.6\%$ & 1.709 \\
EWMA halflife  & $h$=30 days                         & 1.784 & $+0.003$ & 1.984 & $-7.3\%$ & 2.096 \\
               & $h$=120 days                        & 1.607 & $-0.174$ & 1.727 & $-9.1\%$ & 1.538 \\
\bottomrule
\end{tabular}
\smallskip
\caption{\normalfont Hyperparameter sensitivity of FADE. Each row varies one construction parameter from the headline configuration ($K_{\text{nn}}=20$, $K_{\min}=5$, $h=60$ days) with all other elements of GIDF (model architecture, training schedule, Avellaneda--Lee trading rule, universe) held fixed, and $\Delta$Sharpe is measured relative to the headline. The $K_{\text{nn}}=30$ variant exceeded the GPU memory budget at the fourth training window and is omitted.}
\label{tab:fade-sensitivity}
\end{table}

\begin{figure}[!ht]
\centering
\includegraphics[width=1.02\columnwidth]{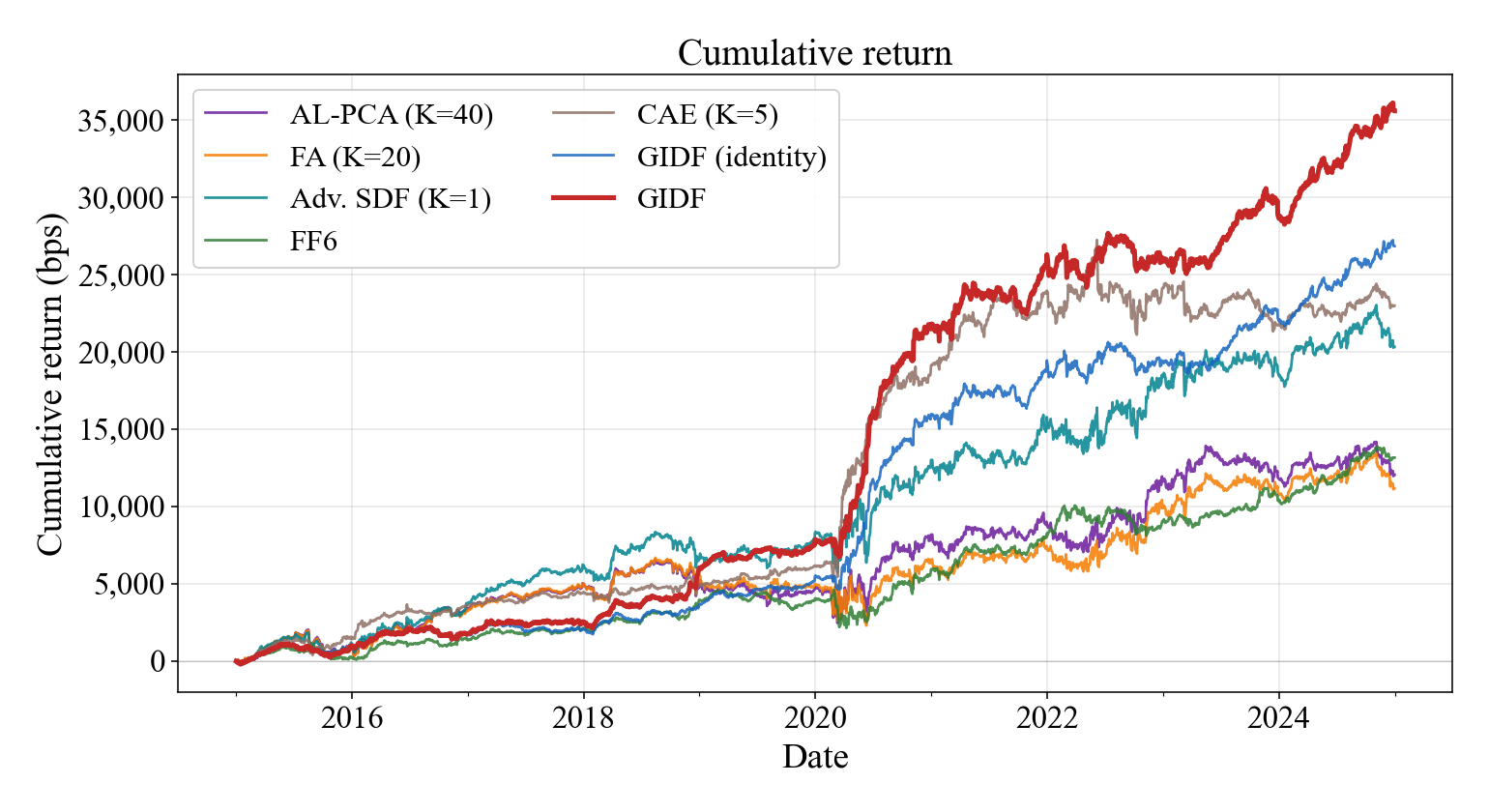}
\caption{\normalfont Cumulative return in basis points across the stitched out-of-sample horizon (2015--2024). GIDF separates from the classical baselines after 2019 and continues to widen the gap through the 2020 COVID episode, the 2022 rate-hike cycle, and the 2023 banking-stress episode.}
\Description{Line chart with dates from 2015 to 2024 on the x-axis and cumulative return in basis points from 0 to 35,000 on the y-axis. Seven curves are plotted: AL-PCA at K equals 40 in purple, FA at K equals 20 in orange, Adversarial SDF at K equals 1 in teal, FF6 in green, CAE at K equals 5 in brown, GIDF identity in blue, and GIDF in red drawn as the thickest line. Between 2015 and early 2020 all seven curves stay bunched together between 0 and 8,000 basis points with only mild separation. In March 2020 there is a brief downward dip across all curves. Immediately after, the GIDF red line jumps sharply upward and separates from the pack. From 2021 onward GIDF continues to climb steadily, reaching about 36,000 basis points by end of 2024. GIDF identity in blue tracks GIDF closely but lower, ending near 27,000. CAE in brown reaches about 23,000. Adversarial SDF in teal reaches about 20,000. AL-PCA purple, FA orange, and FF6 green cluster at the bottom, all ending between 11,000 and 14,000 basis points.}
\label{fig:cumret}
\end{figure}

\vspace{-1mm}
\paragraph{Cumulative return performance.} Figure~\ref{fig:cumret} shows the out-of-sample cumulative raw return of each residual portfolio. Through the first few years the methods track closely and GIDF is not yet clearly ahead. 
The early similarity in performance may reflect differences in the prevailing market environment or the amount of historical data available for training the neural network. Since FADE uses a fixed-length rolling estimation window, however, the nominal sample size of its daily dependency estimator does not increase with the expanding model-training history.
The sharp break comes with the March 2020 COVID dislocation: GIDF ends the crash period positive while every classical baseline loses, and its lead continues to widen through the post-COVID recovery. The 2022 monetary-tightening cycle and the March 2023 regional-bank failures temporarily reverse this direction: classical residuals gain some ground on GIDF during both episodes, but the earlier lead is large enough that GIDF's cumulative curve stays above every classical baseline throughout, and the gap re-widens through the post-2023 recovery. The most striking pattern in the plot is not the widening during shocks but the persistence between them. Even in the calm periods separating each stress episode, the classical baselines never meaningfully close the gap on GIDF: no classical curve crosses GIDF's after the COVID break (the closest approach, AL-PCA in April 2020, remains roughly $0.35$ below GIDF on a $1$-normalized wealth curve), and at the end of the out-of-sample window GIDF's cumulative return is $+356$\% against a best-classical $+132$\%. 
Once established, the cumulative performance advantage persists through the end of the evaluation period, spanning episodes of pandemic-related dislocation, monetary tightening, and sector-specific credit stress, as well as the intervening periods. Although the magnitude of relative outperformance varies across these episodes, the cumulative lead is maintained. These results are consistent with graph-conditioned loadings providing incremental information for statistical arbitrage over the full evaluation period. However, persistence of the cumulative advantage does not necessarily imply uniform outperformance across individual market regimes.

\begin{table}[!ht]
\centering
\small
\begin{tabular*}{\columnwidth}{@{\extracolsep{\fill}}lrrrrr@{}}
\toprule
Entry & GIDF & GIDF (identity) & $\Delta$Sharpe & HAC-$\delta$ $p$ & boot $p$ \\
\midrule
$|s|=1.00$ & 1.921 & 1.715 & $+0.206$ & 0.013$^{**}$ & 0.017$^{**}$ \\
$|s|=1.25$ & 1.781 & 1.593 & $+0.188$ & 0.040$^{**}$ & 0.034$^{**}$ \\
$|s|=1.50$ & 1.479 & 1.314 & $+0.165$ & 0.077$^{*}$\phantom{$^*$} & 0.069$^{*}$\phantom{$^*$} \\
$|s|=1.75$ & 1.336 & 1.157 & $+0.179$ & 0.064$^{*}$\phantom{$^*$} & 0.057$^{*}$\phantom{$^*$} \\
$|s|=2.00$ & 1.023 & 0.759 & $+0.264$ & 0.005$^{***}$ & 0.004$^{***}$ \\
\bottomrule
\end{tabular*}
\smallskip
\caption{\normalfont Sharpe-difference tests of GIDF vs GIDF (identity). Heteroscedasticity- and autocorrelation-consistent (HAC) delta-method~\cite{LedoitWolf2008} and studentized circular block bootstrap ($B=2000$) $p$-values. $^*p<0.10$, $^{**}p<0.05$, $^{***}p<0.01$.}
\label{tab:al-sweep}
\end{table}

\paragraph{Robustness across trading-rule variations.} To confirm that GIDF's edge is not tied to a specific entry threshold, we repeat the ablation of Table~\ref{tab:main} at all five canonical entry thresholds spanning $|s|=1.00$ to $2.00$. Table~\ref{tab:al-sweep} reports the stitched out-of-sample Sharpe ratios for GIDF and its identity ablation, along with delta-method and bootstrap $p$-values for the Sharpe-difference test. The Sharpe gap is positive at every threshold, ranging from $+0.165$ to $+0.264$. It reaches $p<0.05$ at three of five thresholds and $p<0.10$ at all five. Delta-method and bootstrap $p$-values agree to within $\pm 0.01$ at every threshold, so the significance conclusion is not an artifact of the inference procedure. The gap is largest at the tightest threshold ($|s|=2.00$, $\Delta=+0.264$, $p=0.005$), where only the strongest residual dislocations trigger trades. It remains significant at looser thresholds where a larger fraction of the residual distribution is traded. GIDF therefore consistently outperforms GIDF (identity) across the full canonical range of trading-rule cutoffs.

\begin{figure}[!ht]
\centering
\includegraphics[width=1.02\columnwidth]{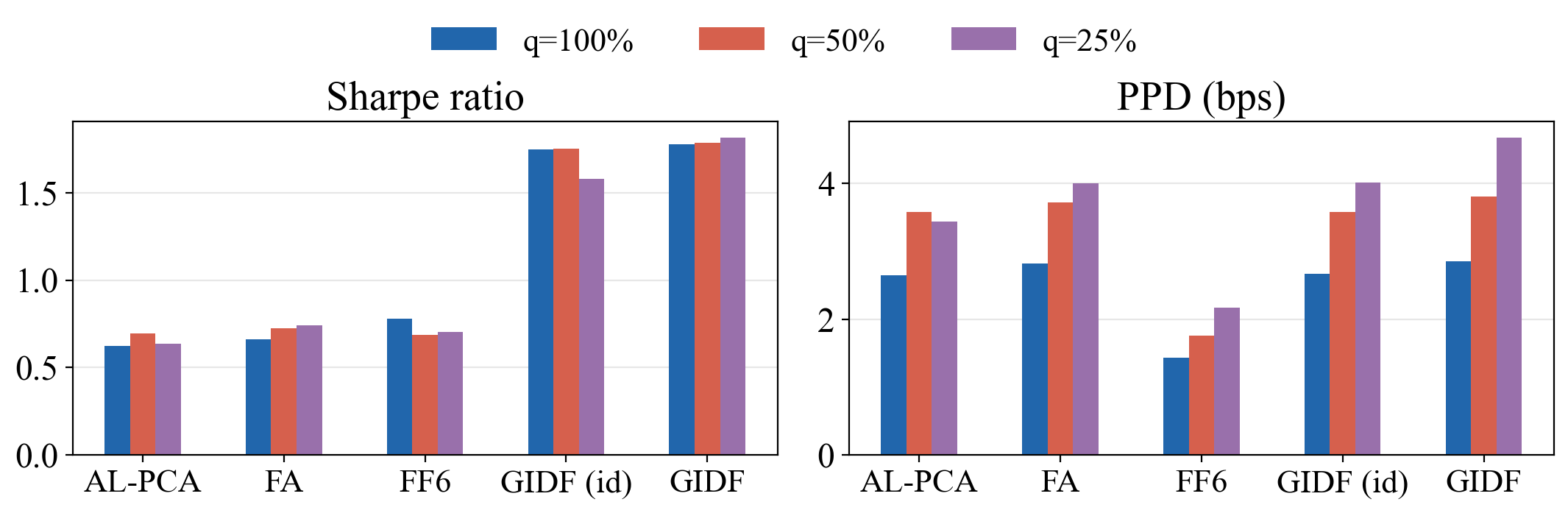}
\caption{\normalfont Quantile-portfolio Sharpe and per-day profit (PPD, bp) for five signals at $q{=}100\%, 50\%, 25\%$. GIDF FADE leads on Sharpe at every bucket; PPD increases as the portfolio becomes more concentrated, indicating that the highest-confidence signals generate the largest profits per position.}
\Description{Two-panel grouped bar chart. Left panel titled Sharpe ratio has y-axis from 0 to about 1.85 and five signal groups on the x-axis: AL-PCA, FA, FF6, GIDF identity, and GIDF FADE. Each group contains three bars in steel blue, muted coral, and muted purple representing quantile buckets q equals 100 percent, q equals 50 percent, and q equals 25 percent. The three classical baselines show Sharpe values between about 0.55 and 0.78, with FF6 slightly higher than AL-PCA and FA. GIDF identity and GIDF FADE both show Sharpe values between about 1.60 and 1.82, dramatically higher than the classical baselines. GIDF FADE is slightly higher than GIDF identity across all three quantiles. Right panel titled PPD in basis points has y-axis from 0 to about 4.75 and the same five signal groups with the same three-quantile bars. AL-PCA, FA, GIDF identity, and GIDF FADE all show PPD between about 2.6 and 4.75, with GIDF FADE reaching the highest values, nearly 4.75 basis points at q equals 25 percent. FF6 has the lowest PPD across all quantiles at 1.4 to 2.2 basis points.}
\label{fig:quantile-bars}
\end{figure}

\paragraph{Robustness to the execution rule.} 
The previous subsection varied the entry threshold within the Avellaneda--Lee rule. Here, we evaluate GIDF under an alternative execution framework. 
We feed the same GIDF s-scores to  AlphaMark~\cite{alphaMark}, an independent benchmark that keeps the signal but replaces the trading rule: it forms unit-bet long--short portfolios by ranking stocks on s-score magnitude and trading the top $q\%$ each day. Since the Avellaneda--Lee book expands when many stocks dislocate and empties when few do, whereas AlphaMark holds fixed breadth every day, agreement across the two rules isolates signal quality from allocation. Figure~\ref{fig:quantile-bars} reports metrics for five signals at $q=100\%$, $50\%$, $25\%$. These analyses compare GIDF with the three classical baselines and the identity ablation; its advantage over the deep conditional models is already established in Table~\ref{tab:main} under the headline rule. At $q=100\%$, GIDF attains Sharpe $1.78$, essentially matching the Avellaneda--Lee headline of $1.781$, suggesting that the performance is driven primarily by the residual signal rather than the execution rule.
GIDF beats every classical baseline in every quantile bucket (Sharpe $1.78$--$1.82$ versus $0.62$--$0.78$), and per-day profit (PPD) rises as the portfolio concentrates on higher-magnitude signals, indicating that stronger GIDF signals are associated with higher per-position profitability.

\begin{table}[!ht]
\centering
\small
\begin{tabular*}{\columnwidth}{@{\extracolsep{\fill}}lrrrr@{}}
\toprule
Benchmark & $\alpha$ (\%/mo) & $t(\alpha)$ & $R^2$ (\%) & FN-SR \\
\midrule
CAPM & 1.862 & 3.84 & 0.23 & 1.253 \\
FF3 & 1.851 & 3.80 & 0.50 & 1.248 \\
FF5~\cite{FamaFrench2015} & 1.801 & 3.72 & 0.76 & 1.216 \\
FF6~\cite{FamaFrench2015,Carhart1997} & 1.800 & 3.71 & 0.77 & 1.215 \\
\bottomrule
\end{tabular*}
\smallskip
\caption{\normalfont Time-series regression of GIDF's D10$-$D1 daily long--short return on classical factor benchmarks (OOS 2015--2024). $\alpha$ and Newey--West $t$-statistic are in monthly percent. $R^2$ is the benchmark's variance share and FN-SR the annualized residual Sharpe.}
\label{tab:decile}
\end{table}

\paragraph{Long-short alpha.} We further evaluate GIDF's per-stock predicted return as a standalone cross-sectional signal.
Each day we sort the tradable universe into deciles by GIDF's predicted return and form a decile long--short portfolio that goes long the top decile and short the bottom. This construction sits in the broader arbitrage-portfolio tradition~\cite{KimKorajczykNeuhierl2021}. The portfolio delivers an annualized out-of-sample return of $22.9\%$ at a Sharpe of $1.28$. Table~\ref{tab:decile} regresses this daily return on four factor-benchmark specifications of increasing richness: the CAPM, the Fama--French three-factor model (FF3), the Fama--French five-factor model (FF5), and the Fama--French six-factor model (FF6). The intercept, measured in monthly percent, remains near $1.80\%$ per month across all four specifications, with Newey--West $t$-statistics from $3.71$ (FF6) to $3.84$ (CAPM) — all above the $t > 3$ threshold advocated by Harvey, Liu, and Zhu~\cite{Harvey2016} for asset-pricing factors. The benchmark $R^2$ stays below $1\%$ in every specification, and the factor-neutral Sharpe, which is the annualized Sharpe of the residual after regressing out benchmark exposure, declines only marginally from $1.28$ to $1.22$. 
The low regression $R^2$ indicates that the selected classical factors explain little of the realized time-series variation in the decile portfolio returns over the evaluation period. Together with the positive estimated intercept, this provides evidence of incremental performance relative to these particular linear factor benchmarks. It does not exclude exposure to omitted, nonlinear, conditional, or time-varying sources of systematic risk.

\begin{figure}[!ht]
\centering
\includegraphics[width=\columnwidth]{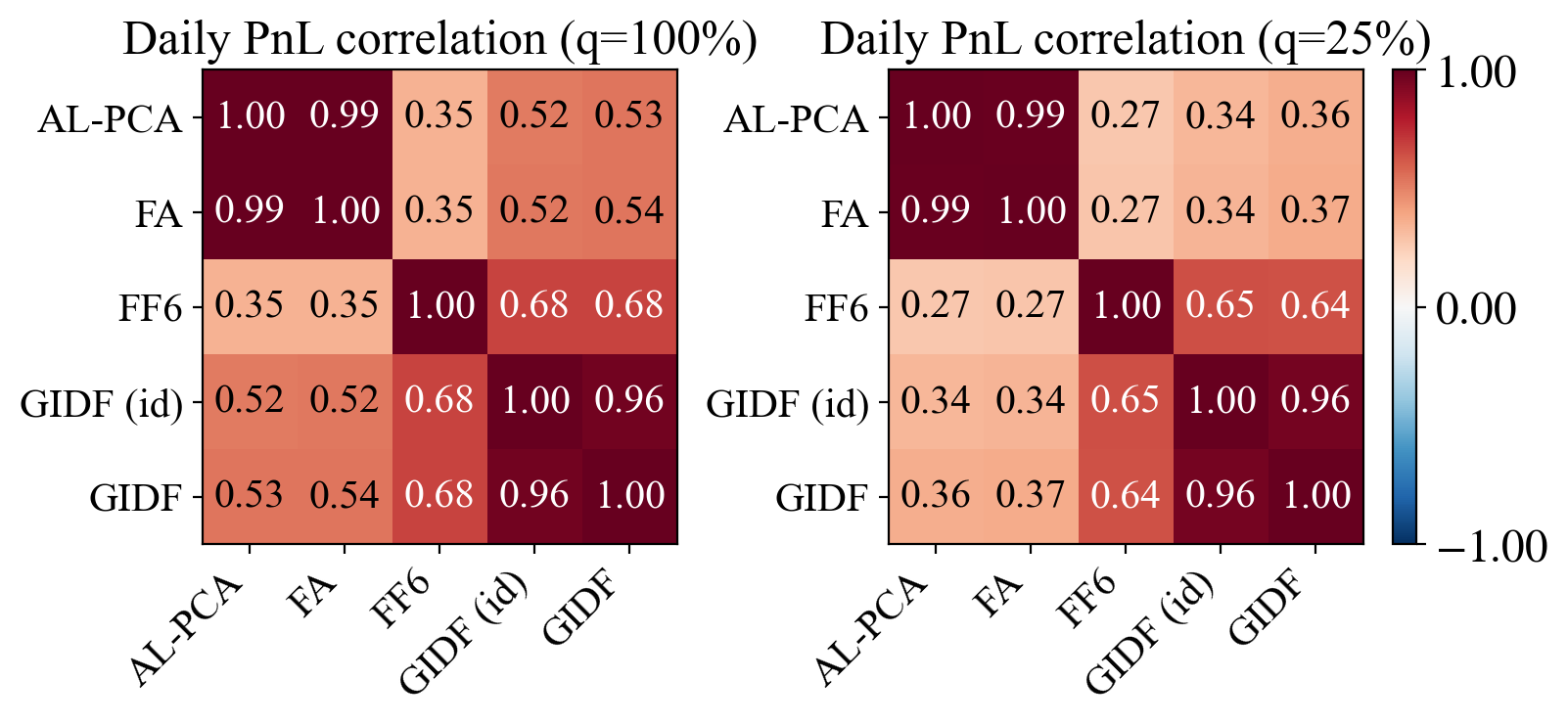}
\caption{\normalfont Spearman correlation of daily long--short PnL across the five signals, at $q{=}100\%$ and $q{=}25\%$.}
\Description{Two five-by-five heatmaps placed side by side, sharing a single vertical colorbar on the right spanning minus 1.00 through 0.00 to plus 1.00 on a red-to-blue diverging scale. Both panels use the same five signals on both axes: AL-PCA, FA, FF6, GIDF identity, and GIDF FADE. Numerical correlation values are printed inside each cell. Left panel titled Daily PnL correlation at q equals 100 percent. Diagonal values are all 1.00. AL-PCA and FA correlate at 0.99. GIDF identity and GIDF FADE correlate at 0.96. GIDF FADE correlates with AL-PCA at 0.53, FA at 0.54, and FF6 at 0.68. FF6 correlates with AL-PCA and FA at 0.35. Right panel titled Daily PnL correlation at q equals 25 percent. Diagonals still 1.00, and the top-tier correlations (AL-PCA and FA at 0.99, GIDF FADE and GIDF identity at 0.96) are unchanged. But GIDF FADE's correlation with AL-PCA drops to 0.36, with FA to 0.37, and with FF6 to 0.64. FF6 correlations with AL-PCA and FA drop from 0.35 to 0.27. All values are positive throughout both panels.}
\label{fig:alphamark-h2}
\end{figure}

\paragraph{Correlation with classical residual signals.} Figure~\ref{fig:alphamark-h2} reports the Spearman correlation of daily long--short PnL across the five signals, for the full ($q{=}100\%$) and concentrated ($q{=}25\%$) portfolios. GIDF's daily PnL correlates moderately with all three classical baselines and most strongly with the Fama--French residual ($0.68$ at $q{=}100\%$, $0.64$ at $q{=}25\%$). As the portfolio concentrates on its strongest signals, correlation with the PCA and factor-analysis PnL falls sharply ($0.53$ to $0.36$) while correlation with Fama--French remains comparatively stable. GIDF's highest-conviction trades therefore co-move with the economic Fama--French factors on the same days but drift away from the statistical PCA and factor-analysis baselines. Daily PnL co-movement is not linear spanning. Table~\ref{tab:decile} tests the stronger question directly: regressing GIDF's decile long--short return on the Fama--French six-factor model yields $R^2 < 1\%$. GIDF's cross-sectional return spread is therefore largely unspanned by Fama--French factor exposures, even though the trading books co-move day-to-day. GIDF trades within the same residual-arbitrage universe as the classical benchmarks while carrying pricing information not well captured by their linear factor structures.

\begin{table}[H]
\centering
\small
\begin{tabular*}{\columnwidth}{@{\extracolsep{\fill}}lrrrr@{}}
\toprule
Factor model               & Plain A\&L      & + SPONGE cluster-$\sigma$ & $\Delta$ \\
\midrule
AL-PCA ($K=40$)            & $0.550$         & $0.574$          & $+0.024$ \\
FA ($K=20$)                & $0.524$         & $0.641$          & $+0.116$ \\
FF6                        & $0.930$         & $0.962$          & $+0.033$ \\
GIDF (identity)            & $1.593$         & $1.783$          & $+0.190$ \\
\textbf{GIDF ($K=1$)}      & $1.781$         & $1.875$          & $+0.094$ \\
\bottomrule
\end{tabular*}
\smallskip
\caption{\normalfont SPONGE~\cite{Cucuringu2020} clusters from FADE compute per-cluster residual $\sigma$ inside the Avellaneda--Lee rule. Sharpe rises for every method versus strict own-stock $\sigma$.}
\label{tab:sponge}
\end{table}

The full GIDF model and its identity-graph variant exhibit a daily PnL correlation of approximately $0.96$ under both portfolio concentration levels. This indicates that the two specifications share a substantial common return component, with FADE primarily improving the performance of an existing signal rather than necessarily generating an independent source of trading returns.

\vspace{-1mm}
\paragraph{Cluster-conditional $\sigma$ normalization} SPONGE~\cite{Cucuringu2020} is a spectral clustering~\cite{vonLuxburg2007} method for signed graphs. Given a signed adjacency $A$, SPONGE separates it into a positive part $A^{+} = \max(A, 0)$ and a negative part $A^{-} = -\min(A, 0)$ and forms their normalized Laplacian $L^{+}$ and $L^{-}$. It solves the generalized eigen-problem $(L^{+} + \tau I)\, v = \lambda\, (L^{-} + \tau I)\, v$ with a small regularization $\tau > 0$. The leading eigenvectors are $k$-means clustered into $K$ groups so that positively-linked stocks fall inside the same cluster and negatively-linked stocks fall in different clusters. We feed SPONGE with the FADE graph. We average the per-day FADE signed normalized Laplacian $\hat L_t$ over the training window into a single static Laplacian $\bar L = T_{\text{train}}^{-1} \sum_t \hat L_t$ and recover the corresponding signed adjacency $A = I - \bar L$. SPONGE runs on $A$ with $K = 10$ clusters and $\tau = 0.1$. The partition is balanced (largest cluster $24\%$ of stocks) and is fixed for the out-of-sample period. Each stock's Avellaneda--Lee $s$-score is normalized by its cluster's residual standard deviation instead of its own-stock $\sigma$. Table~\ref{tab:sponge} shows that this cluster-$\sigma$ augmentation lifts the Sharpe of every returns-based factor model in the table. GIDF (FADE) reaches Sharpe $1.875$, a further $+0.094$ above the strict-AL headline of $1.781$. The same factor-stripping mechanism that gives FADE its balanced signed edges also yields informative SPONGE clusters.
These results suggest that graph-based clustering provides a useful cross-sectional structure for residual normalization and signal allocation. The performance improvement may reflect more stable volatility estimates, improved comparability of signals across related assets, or more effective position selection, rather than a change in the underlying residual mean-reversion dynamics.

\begin{table}[!h]
\centering
\small
\begin{tabular*}{\columnwidth}{@{\extracolsep{\fill}}lrrrrr@{}}
\toprule
Model & Turnover & Gross & 1\,bp & 2\,bp & 5\,bp \\
\midrule
AL-PCA ($K=40$)             & 180\% & 0.550 & 0.285 & 0.020 & $-0.776$ \\
FA ($K=20$)                 & 181\% & 0.524 & 0.259 & $-0.006$ & $-0.801$ \\
FF6                           & 155\% & 0.930 & 0.521 & 0.112 & $-1.115$ \\
\midrule
GIDF (identity)               & 152\% & 1.593 & 1.141 & 0.687 & $-0.679$ \\
\textbf{GIDF ($K=1$)}       & \textbf{154\%} & \textbf{1.781} & \textbf{1.341} & \textbf{0.900} & \textbf{-0.426} \\
\bottomrule
\end{tabular*}
\smallskip
\caption{\normalfont Transaction-cost sensitivity. Turnover is the average daily fraction of positions traded; net Sharpe is shown for $1$, $2$, and $5$\,bps per unit traded. GIDF has the highest net Sharpe at every cost level.}
\label{tab:cost-sensitivity}
\end{table}

\paragraph{Sensitivity to transaction costs.} Table~\ref{tab:main} reports gross performance. Table~\ref{tab:cost-sensitivity} recomputes the net Sharpe under costs of $1$, $2$, and $5$ basis points per unit traded~\cite{NovyMarxVelikov2016}. 
Daily turnover is similar across methods ($152$--$181\%$), and GIDF is among the lowest ($154\%$), below the PCA and factor-analysis baselines, so its large gross margin does not come from faster trading. Both GIDF variants are thus structurally cost-efficient before any bps are netted out. GIDF has the highest net Sharpe at every cost level. It remains clearly profitable through $2$ bps (Sharpe $1.34$ at $1$ bp and $0.90$ at $2$ bp), where the best classical baseline, Fama--French, is already near zero ($0.52$ and $0.11$). For large-cap U.S. equities, institutional execution costs are often estimated in the low-single-digit basis-point range, so the positive Sharpe at $2$ bps indicates that the result survives economically meaningful trading frictions. 
At $5$ bps all five strategies are unprofitable, but GIDF degrades the least ($-0.43$, versus $-0.68$ to $-1.12$). GIDF's lead over every baseline therefore persists once trading frictions are included, not only in the gross figures.

\section{Conclusion}
\label{sec:conclusion}
The results indicate that the cross-sectional dependency structure between stocks, when isolated from systematic factor exposure and modeled with signed edges, contains economically meaningful information about daily residual mean-reversion. 
Controlled ablations show that introducing the FADE graph improves GIDF relative to its identity-graph variant, while the full GIDF model also outperforms classical and deep conditional factor benchmarks, with the advantage persisting across both calm and stressed regimes. 
When deployed as a cross-sectional long--short signal, GIDF delivers positive estimated alpha relative to the selected classical linear factor benchmarks. These findings provide evidence that graph-conditioned return decompositions contain economically useful information complementary to conventional factor specifications, although further analysis is required to establish their incremental contribution relative to broader risk models and existing statistical-arbitrage signals.

These findings suggest that incorporating cross-sectional dependency structure into the factor-loading stage provides information complementary to firm characteristics alone, rather than simply improving the downstream execution policy.

The framework admits two natural extensions. First, GIDF uses a single-factor architecture at $K=1$. Extending it to multiple factors decomposes systematic co-movement across $K$ dimensions instead of compressing it into one. Each factor could learn its own graph-conditioned loading pattern and potentially capture a distinct source of cross-sectional dependency. 
A further extension is to distinguish factor-mimicking portfolio weights from return-factor loadings, allowing the pricing and reconstruction components to operate through related but non-identical representations. The framework also admits alternative graph constructions based on economically specified factor residuals, directed lead--lag relationships, order-flow dependencies, and multiple observation frequencies. Such extensions could help distinguish persistent economic relationships from transient cross-sectional trading pressure.
Second, we decoupled residual construction from execution to isolate the graph's loading-stage contribution, not because of any framework limitation. A richer execution scheme, such as a graph-attention trading policy on the same FADE Laplacian or end-to-end optimization of net-of-cost performance, can be applied layer on without altering the residual model.
\bibliographystyle{plainnat}
\bibliography{references}

\end{document}